\documentclass[aps,pra,reprint,groupedaddress,amssymb]{revtex4-1}
\usepackage{natbib}
\usepackage{bm}
\usepackage{textcase}
\usepackage{hyperref}
\usepackage{graphicx}
\usepackage{color}
\usepackage{xspace}
\usepackage{cancel}
\newcommand{\ket}[1]{\ensuremath{{\left|#1\right\rangle}}\xspace}

\begin{document}

\title{Microwave preparation of two-dimensional fermionic spin mixtures}

\author{B. Peaudecerf}\email{bruno.peaudecerf@strath.ac.uk}
\author{M. Andia}
\author{M. Brown}
\author{E. Haller}
\author{S. Kuhr}\email{stefan.kuhr@strath.ac.uk}
\affiliation{University of Strathclyde, Department of Physics, 107 Rottenrow East, Glasgow G4 0NG}

\date{\today}

\begin{abstract}
We present a method for preparing a single two-dimensional sample of a two-spin mixture of fermionic potassium in a single antinode of an optical lattice, in a quantum-gas microscope apparatus. Our technique relies on spatially-selective microwave transitions in a magnetic field gradient. Adiabatic transfer pulses were optimized for high efficiency and minimal atom loss and heating due to spin-changing collisions. We have measured the dynamics of those loss processes, which are more pronounced in the presence of a spin mixture. As the efficient preparation of atoms in a single antinode requires a homogeneous transverse magnetic field, we developed a method to image and minimize the magnetic field gradients in the focal plane of the microscope.

\vspace{1cm}
\end{abstract}

\maketitle


\section{\label{Intro}Introduction}

The preparation and study of states of ultracold matter with low dimensionality is a fruitful area of research: these systems can exhibit unique properties, such as the Berezinskii-Kosterlitz-Thouless (BKT) transition for a 2D gas \cite{Clade2009,Hadzibabic2006}; fermionic atoms in a 2D optical lattice realize the 2D Hubbard model, which is thought to contain the key mechanism to high-$T_c$ superconductivity \cite{Lee2006}. The past few years have seen  in particular the rise of quantum-gas microscope experiments with fermionic species as a tool for the study of 2D lattice physics, with single-atom and single-lattice-site resolution  \cite{Haller2015,Omran2015,Parsons2015,Cheuk2015,Edge2015}. This has led to remarkable achievements such as the observation of band- and Mott-insulator phases of fermions \cite{Omran2015,Greif2016,Cheuk2016PRL}, the measurement of charge and spin correlations \cite{Cheuk2016,Parsons2016}, the realization of an antiferromagnetic phase \cite{Mazurenko2017,Boll2016,Brown2017}, or the study of many-body localization \cite{Choi2016}.

For quantum-gas microscope experiments it is a particular requirement that one prepare a single, two-dimensional atomic sample in the focal plane of an optical microscope. This demanding preparation can rely on a variety of techniques: they can combine magnetic and optical traps \cite{Gillen2009} to load a single antinode of a one-dimensional optical lattice; make use of blue-detuned, repulsive optical traps created by a spatial light modulator \cite{Rath2010} or by using Laguerre-Gauss beams \cite{Dyke2016}. Other methods use an accordion optical lattice with adjustable spacing \cite{Williams2008,Ville2017}, which allows for the compression of an initially large atomic sample into a single 2D system; or employ spatially-dependent microwave transitions in a magnetic gradient \cite{Sherson2010,Haller2015} to isolate a single plane of a short-wavelength optical lattice.

Here we report on such a microwave-based preparation technique that allows us to prepare a mixture of two magnetic states of fermionic potassium in a single antinode of a red-detuned optical lattice, in our \mbox{quantum-gas} microscope apparatus.
The procedure relies on a sequence of adiabatic microwave transfer pulses \cite{Khudaverdyan2005} in a
magnetic field gradient, which are optimized for efficiency and speed.
Compared to the technique used in our previous publication \cite{Haller2015}, the scheme presented here introduces several new features, such as the simultaneous use of two spin states, the visualization of magnetic field gradients and the use of microwave transfers instead of optical pulses, whenever possible. We developed this scheme in order to minimize heating induced by photon scattering in the previously used technique \cite{Haller2015}, with the goal to increase the phase-space density, essential for the preparation of strongly-correlated fermionic phases. Using two spin states makes the scheme significantly more involved, because we have to use several microwave pulses at different frequencies that need to be matched to the same antinode. When optimizing our scheme, we observed fast inelastic collisions that only arise when the atoms are in different spin states of the upper hyperfine manifold of the ground state. The frequency sensitivity of the pulses also allowed us to perform spectral imaging \cite{Bohi2010, Marti2018} and visualize our magnetic field gradients directly on a fluorescence image.

The paper is organized as follows: in Section \ref{sec:preparation} we describe the procedure used and demonstrate our ability to prepare a single 2D system containing the two magnetic states. In Section \ref{sec:optimization} we detail the characteristics of the adiabatic microwave transfer pulses employed, and the constraints they have to satisfy. In particular, we present measurements of the loss dynamics in the upper hyperfine manifold of the ground state which limit the duration of the pulses. Finally, in Section \ref{sec:gradients} we describe how we can use the sharp spectral features of our preparation method to image magnetic field gradients in our system.

\section{\label{sec:preparation}Selective preparation in a lattice antinode}

\subsection{Experimental setup}

Our quantum-gas microscope apparatus is designed to study fermionic potassium ($^{40}$K) atoms in optical lattice potentials (details of our setup can be found in an earlier publication \cite{Haller2015}).
Starting with a gas at room temperature, several steps are necessary to cool the atoms down to quantum degeneracy and to prepare a two-dimensional (2D) layer of atoms close to the microscope objective.
In our setup, atoms are first captured in a 2D magneto-optical trap (2D-MOT), from which they are sent towards a 3D-MOT. After a phase of two-photon Raman gray molasses cooling on the D$_2$ line \cite{Bruce2017}, the atoms are loaded into a red-detuned crossed optical dipole trap which captures $9.2(3)\times 10^6$ atoms. Then, atoms are transported to another section of the vacuum chamber above the microscope by displacing the focal point of a red-detuned optical dipole trap beam. 
The atoms are subsequently loaded into a second red-detuned, crossed optical dipole trap, are prepared in a mixture of two magnetic sub-levels (see Section \ref{Procedure}) and are cooled down to quantum degeneracy by forced evaporation. We reach a temperature of $T=45(10)$\,nK for $3.1(1)\times 10^4$ atoms which corresponds to a fraction $T/T_F=0.18(2)$ of the Fermi temperature $T_F$.
The next experimental step, subject of the study in this paper, consists in the preparation of a quantum gas in a single 2D antinode, or "layer" of atoms, in the focal plane of the microscope objective. In our scheme, many layers are initially populated and we remove atoms from all but one layer \cite{Sherson2010}.

\begin{figure}[ht!]
\includegraphics[width=0.9\linewidth]{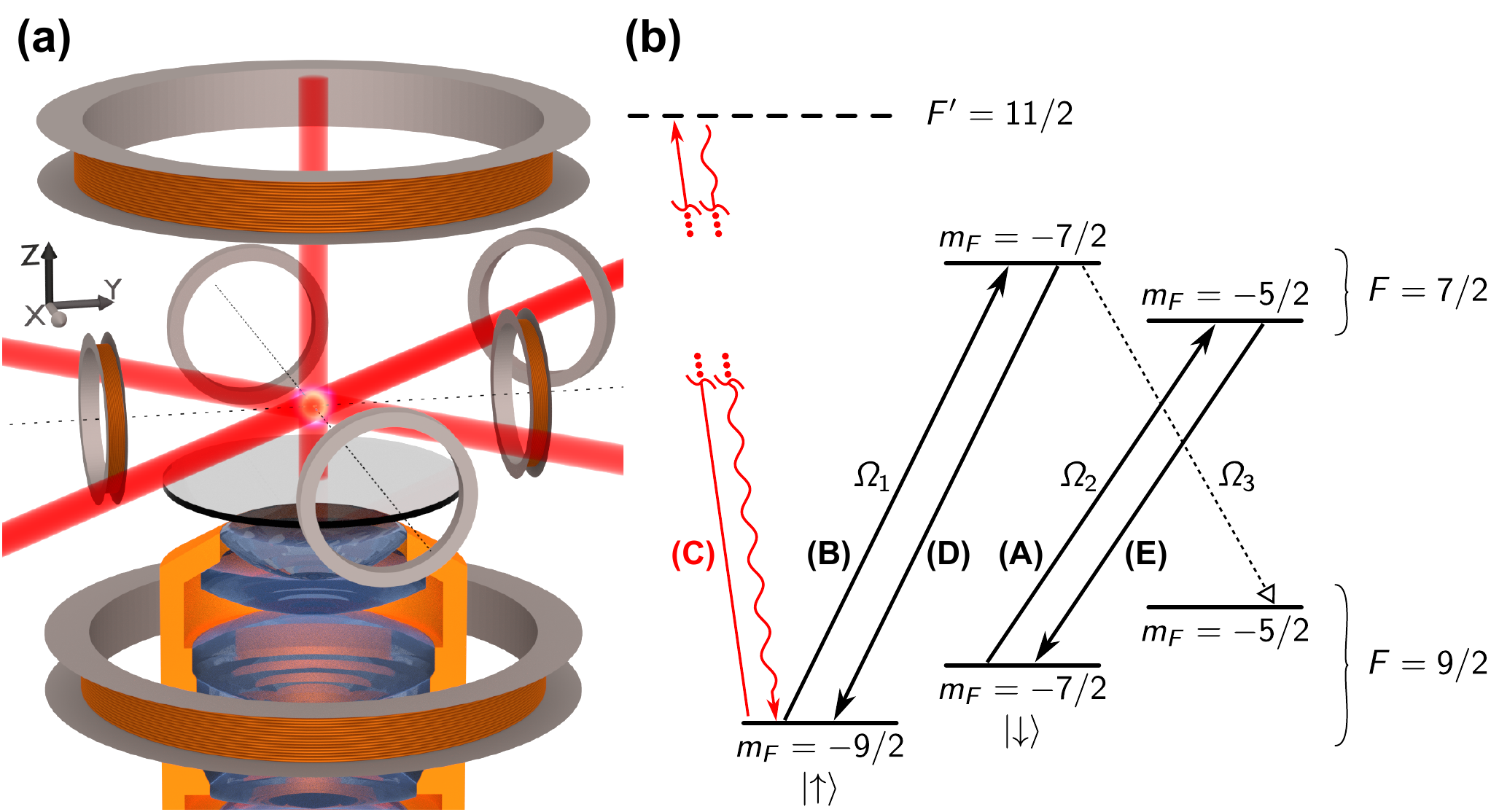}
\caption{\textbf{(a)} Illustration of our experimental setup, showing the microscope objective (bottom) and the magnetic coils used to control the magnetic field at the level of the atoms. \textbf{(b)} Simplified level scheme of the $^2$S$_{1/2}$ state for $^{40}$K \color{black} atoms in presence of a  magnetic field. Circularly polarized microwave transitions 1 to 3 couple magnetic sublevels in the $F=9/2$ and $F=7/2$ manifolds with Rabi frequencies $\Omega_1$ to $\Omega_3$. Our layer preparation scheme with steps A to E is described in the text.}
\label{setup}
\end{figure}

The lattice potential is formed by a vertical laser beam, which is retro-reflected off a vacuum window close to the microscope objective (figure \,\ref{setup}(a)). The wavelength and the waist of the beam are respectively $\lambda_L=1064\,\mathrm{nm}$ and 110\,$\mathrm{\mu}$m. After transfer from the crossed optical dipole trap to the lattice potential, the atoms occupy approximately $50$ layers with $1.2\times 10^3$ particles in the central layer.

Our preparation scheme of a single lattice layer relies on spatially-selective microwave transitions in a vertical magnetic field gradient and on a spin-state-dependent optical removal process. The gradient is generated by two vertical coils with counter-propagating currents. These coils produce a magnetic quadrupole field with a field minimum a few millimeters above the atoms. Two pairs of shim coils are used to position the field minimum in the horizontal plane. Field gradients in the horizontal plane can be compensated by a fifth horizontal coil as illustrated in figure\,\ref{setup}(a). With this configuration, we generate a gradient of $\partial_z B=7.27\,\mathrm{mG/\mu m}$ and a field strength of $B_0=11.60\,\mathrm{G}$ at the position of the atoms.
The field $B_0$ is mostly due to a vertical distance of about 4\,mm between the center of the quadrupole coils and the atom cloud. The field is therefore essentially vertical and homogeneous horizontally at the level of the atoms (we use our preparation technique to probe residual gradients of the field value in section \ref{sec:gradients}).

\subsection{Microwave transfer scheme}
\label{Procedure}

A sketch of the level structure and transitions involved in the single-layer preparation protocol is presented in figure \,\ref{setup}(b). The atoms are initially in a balanced statistical mixture of the $\ket{m_F=-9/2}\equiv\ket{\uparrow}$ and $\ket{m_F=-7/2}\equiv\ket{\downarrow}$ states of the $F=9/2$ manifold, which is prepared prior to the evaporative cooling in the crossed optical dipole trap. It is this mixture of states that makes our microwave transfer scheme challenging. We use two microwave pulses to drive $\sigma^+$ transitions to the states $\ket{F=7/2,m_F=-7/2}$ (transition 1) and $\ket{F=7/2,m_F=-5/2}$ (transition 2) at the position corresponding to the same layer. Both transitions show a position-dependent frequency shift due to the Zeeman effect, and our values of $B_0$  and $\partial_z B$ result in a frequency shift of 7\,MHz between the transitions, and a spatially-dependent frequency shift for transition 1 (2) of $\Delta_1=9.68$\,kHz ($\Delta_2 = 7.37$\,kHz) between adjacent lattice layers (separated by 532\,nm). We use microwave pulses with a spectral width of 5.25\,kHz and 7.0\,kHz to selectively address atoms in a specific spin state in a particular layer, while minimizing unwanted transfer of atoms at other positions. In the following, we denote $\mathcal{L}_0$ the vertical lattice layer located at the focus of our microscope, in which we want to prepare our atomic sample; other layers are denoted $\mathcal{L}_n$, $n\neq 0$.

The detailed protocol involves five steps, labeled  A to E  in figure \ref{setup}(b). A first microwave pulse A with an efficiency of 98\% (see section \ref{sec:optimization}) drives transition 2 resonantly only for atoms in state $\ket{\downarrow}$ located in the selected layer $\mathcal{L}_0$ of the vertical lattice, transferring them to the state $\ket{F=7/2,m_F=-5/2}$.  A second microwave pulse B then transfers atoms in state $\ket{\uparrow}$ in the same layer $\mathcal{L}_0$ to state $\ket{F=7/2,m_F=-7/2}$. At this stage, 98\% of the atoms in layer $\mathcal{L}_0$ are in the $F=7/2$ manifold, whereas atoms in other layers remain in the $F=9/2$ manifold. A laser pulse C, tuned to the cycling $F=9/2\rightarrow F'=11/2$ transition of the D$_2$ line, removes all atoms in the $F=9/2$ manifold after a few optical cycles (corresponding to a pulse duration of 0.5\,ms) by heating, effectively emptying all layers other than $\mathcal{L}_0$. Atoms in $\mathcal{L}_0$ are then returned to the $F=9/2$ manifold with a pair of microwave pulses on transitions 1 and 2 (steps D and E). Atoms which were not transferred by these microwave pulses -- around 2\% due to the efficiency of the microwave transfer -- are then returned to the $F=9/2$ manifold with a short optical re-pumping pulse tuned to the $F=7/2\rightarrow F'=9/2$ transition.

Due to off-resonant scattering from the $F'=9/2$ manifold during the removal pulse C, atoms can 
also decay to the $F=7/2$ manifold of the ground state by spontaneous emission, such that they become unaffected by the removal beam. 
This leads to a fraction of atoms surviving the removal pulse in layers other than $\mathcal{L}_0$ of 0.3(1)\%. These atoms, located out of the focal plane of the microscope, would appear in a fluorescence image as a diffuse background signal, reducing the visibility of atoms in the desired layer $\mathcal{L}_0$. Repeating steps A to E a second time
allows us to reach high efficiencies in removing these atoms located in layers other than $\mathcal{L}_0$, ensuring a negligible background signal for single-atom fluorescence images.

\subsection{Experimental demonstration of the layer preparation}
\label{Realization}

\begin{figure}
\includegraphics[width=\linewidth]{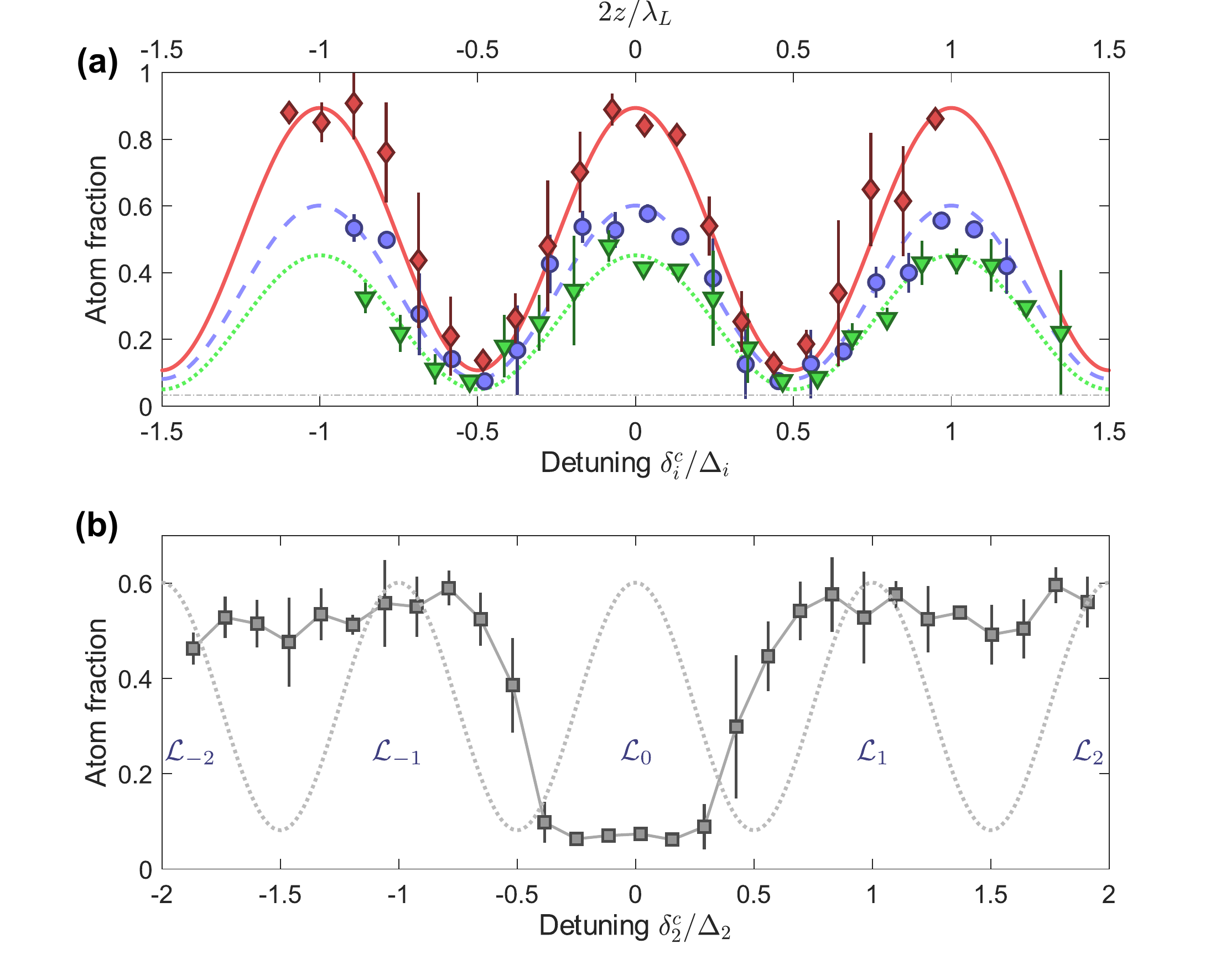}
\caption{\textbf{(a)} Sample preparation in successive layers of an optical lattice. We measure the atom population by fluorescence imaging after addressing only atoms in state $\ket{\uparrow}$ (dashed line with blue circles), in $\ket{\downarrow}$ (dotted line with green triangles) or both states (solid line with red diamonds) at a given height $z$, corresponding to a certain microwave detuning $\delta_i^c$ ($i=1,2$). The observed peaks correspond to the centers of successive antinodes of the vertical lattice. Each detuning $\delta_i^c$ is rescaled by the corresponding frequency shift per lattice layer $\Delta_i$. The small offset of 3.2(1)\% measured independently (dash-dotted gray line) arises from the few atoms not removed by the optical pulse.  \textbf{(b)} Matching the addressing frequencies to a single layer $\mathcal{L}_0$. The position of neighboring layers $\mathcal{L}_n$, $n\neq 0$, is also represented for reference. The measured atom population is normalized to the total atom number in a single layer, extracted from the peak values of the blue/dashed and green/dotted fits in (a).}
	\label{slicing}
\end{figure}

We first demonstrate our layer preparation technique for only one spin state in the mixture. For this purpose we first omitted the microwave pulses A and E and scanned the frequency of the pulses B and D, which selects the atoms in state $\ket{\uparrow}$ in a well-defined layer $\mathcal{L}_0$ of our lattice. All other atoms, including atoms in state $\ket{\downarrow}$ in $\mathcal{L}_0$, are removed by the laser pulse C. The remaining atom number at the selected lattice layer was determined by fluorescence imaging with our microscope objective \cite{Haller2015}.
The blue circles in figure\,\ref{slicing}(a) show the number of remaining atoms for a changing detuning $\delta_1^c$ of the microwave pulses for transition 1, normalized to the total number of atoms present in a single layer of the vertical lattice. We measure the microwave detuning relative to the resonance frequency for atoms in the lattice layer $\mathcal{L}_0$ (the specifics of our microwave pulses will be discussed in Section \ref{MWpulses}). We observe periodic oscillations of the atom number with a period given by the frequency shift per lattice layer $\Delta_1$. The oscillations are caused by successive resonances of the transfer pulses with atoms in adjacent lattice layers. The same measurement can be performed for the spin state $\ket{\downarrow}$ using microwave transitions A and E only (green triangles on figure\,\ref{slicing}(a)). The small offset (dash-dotted line in figure\,\ref{slicing}(a)) is due to background fluorescence from atoms in other layers that have not been removed by the optical pulse and an additional offset of individual curves is due to off-resonant transfer when the microwave center frequency is between two neighboring layers (see section \ref{optimise}).

The data in figure\,\ref{slicing}(a) provides an additional measurement of the population balance in our spin mixture. Simultaneous sine fits yield a spin proportion of 58(1)\% of the atoms in state $\ket{\uparrow}$. These proportions are identical to those measured in the dipole trap prior to loading the vertical lattice for this dataset.

The previous measurement demonstrates the selection of a single lattice layer for each spin state, but it does not verify that atoms in the same layer are addressed by both transitions. In order to prepare a sample containing both spin states at the very same lattice layer, we have to match the frequencies of transitions 1 and 2 to address atoms at the same position. We exploit the fact that transition 2 is degenerate with the $\sigma^-$ transition 3, i.e. $\ket{7/2,-7/2}\rightarrow \ket{9/2,-5/2}$ in figure \,\ref{setup}(b). Starting with a sample of atoms in state $\ket{\uparrow}$ only, we use a first microwave pulse B to transfer all atoms in the selected lattice layer to state $\ket{7/2,-7/2}$. A second microwave pulse with detuning $\delta_2^c$ only addresses these atoms if its frequency matches the resonance frequency for transition 3 at the selected lattice layer.
On resonance, those atoms are transferred to state $\ket{9/2,-5/2}$, and they are subsequently removed by the laser pulse. This pulse addresses all remaining atoms in the $F=9/2$ manifold, both the atoms in state $\ket{m_F=-5/2}$ in $\mathcal{L}_0$ and those in state $\ket{\uparrow}$ in layers $\mathcal{L}_{n\neq 0}$.
As in the previous measurement, we detected the number of remaining atoms by fluorescence imaging.

We observed a minimum of the atom number when the frequencies of microwave pulses for transitions 1 and 3 address the same layer (figure\,\ref{slicing}(b)). For other frequencies, we detected a constant fluorescence signal when the two transitions address atoms in different layers. The dotted line in figure\,\ref{slicing}(b) illustrates the variation of intensity of the red-detuned optical lattice beam,
and the labels $\mathcal{L}_n$ indicate the positions of the lattice antinodes.

With the knowledge of both resonance frequencies for transitions 1 and 2, we are able to prepare a single layer of fermions in a mixture of the $\ket{\uparrow}$ and $\ket{\downarrow}$ states, by following steps A to E as described earlier. To demonstrate this, we scan the center frequencies $\delta_1^c$ and $\delta_2^c$ of the microwave pulses for transitions 1 and 2 in parallel, and measure the number of remaining atoms by fluorescence imaging. We keep the ratio $\delta_2^c/\delta_1^c=3/4$ constant during the measurement to take into account the respective Zeeman shifts of the transitions. Again, we observed a periodic signal as atoms in both spin states in successive lattice layers are addressed resonantly (red diamonds in figure \,\ref{slicing}(a)). The peak value of the fluorescence signal is close to the sum of the peak values observed for the transfer of each individual spin state. This confirms the transfer of all atoms in both spin states within the same lattice layer. We assess the transfer efficiency for individual spin states in the following Section, and we also measure atom loss (see Section \ref{sec:losses}) which explains that the contrast of the fringe for the transfer of both spin states is smaller than expected.
The field offset $B_0=11.60\,\mathrm{G}$ prevents addressing of other layers by the microwave pulses. Any pulse addressing transition 1 (resp. 2) in the layer in the focal plane of the microscope is also resonant with transition 2 (resp. 1) in a layer located $400\,\mathrm{\mu m}$ (resp.~$520\,\mathrm{\mu m}$) away, but these distances are much larger than the vertical sample size of $50\,\mathrm{\mu m}$.

Our all-microwave preparation method is an extension of existing techniques used in previous work \cite{Sherson2010, Haller2015}, which relied on a combination of microwave transfer and optical pumping pulses. The advantage of a protocol that relies solely on microwave pulses is that it does not induce any noticeable heating of the selected atoms due to photon scattering and does not cause transfer of atoms to higher lattice bands. 

\section{\label{sec:optimization}Optimizing the microwave transfer pulses}
\label{optimise}
\subsection{Adiabatic pulses}
\label{MWpulses}

We optimize the amplitude and frequency sweeps during our microwave pulses to reach a maximal transfer efficiency at layer $\mathcal{L}_0$ and a minimal transfer at other layers $\mathcal{L}_{n\neq 0}$. The value of the magnetic field gradient and the spacing of the lattice layers induce a limit to the available frequency range, and we need to restrict our frequency sweeps of the microwave transitions to a few kHz. We use hyperbolic secant pulses (HS1) \cite{Khudaverdyan2005} to drive adiabatic passages \cite{garwood2001} with flat transfer windows and sharp spectral edges.
During an HS1 pulse the frequency detuning $\delta_i(t)$ and the coupling amplitude $\Omega_i(t)$ change according to
\begin{eqnarray}
\delta_i(t)&= &\frac{\delta^0_{i}}{2} \tanh(2t/\tau)+\delta^c_i\\
\Omega_i(t)&= &\Omega_i\, \mathrm{sech}(2t/\tau)
\end{eqnarray}
for $-T/2<t<T/2$, with a characteristic timescale $\tau$. Here, $\Omega_i$ is the Rabi frequency associated with microwave transition $i$ ($i=1,2$), and $\delta^0_{i}$ the frequency width of the pulses.
We ensure a smooth switch-on process of all pulses by using a ratio between $\tau$ and the pulse duration $T$ of $\tau/T=1/5$ \cite{garwood2001}.

The transfer efficiencies depend critically on the choice of $\delta^0_i$ and $T$. The frequency width must be small enough to prevent the addressing of neighboring lattice layers, and large enough to account for small fluctuations of the magnetic field, with frequency drifts of the order of 1\,kHz for transition 1. The pulse duration $T$ must be short enough to avoid atom loss (section \ref{sec:losses}), and sufficiently long to
obey the adiabaticity condition \cite{vitanov2001}:
\begin{eqnarray}
\frac{|\dot{\delta}\Omega-\delta\dot{\Omega}|}{(\delta^2+\Omega^2)^{3/2}} \ll 1.
\end{eqnarray}
We measured a Rabi frequency of $\Omega_{1}=6.1(1)\,\mathrm{kHz}$, and deduce the frequency $\Omega_{2}= \sqrt{7/9}\,\Omega_1=5.4(1)\,\mathrm{kHz}$ using the Clebsch-Gordan coefficients of the transitions.
We performed numerical calculations of the time-dependent dynamics of the effective two-level system coupled by the microwave transition to guide our choice of parameters for the adiabatic transfer pulses.
This lead us to choose frequency widths of $\delta^0_1=7.0\,\mathrm{kHz}$ and $\delta^0_2=5.25\,\mathrm{kHz}$ over a duration $T=1\,\mathrm{ms}$, as a way to maximize stability while keeping the addressing of neighboring lattice layers to a negligible amount. In the following paragraph we characterize the dependence of the transfer pulse on the duration $T$.

\subsection{Experimental characterization of the microwave transfer}

We characterize the transfer efficiency of the microwave pulses starting from an equal mixture of the states $\ket{\uparrow}$ and $\ket{\downarrow}$. For simplicity we optimize the microwave transfer for atoms in a homogeneous magnetic field instead of a field gradient. This allows us to use all atoms held in the vertical lattice, instead of a single layer, which increases the signal-to-noise ratio. The final atom number after the microwave transfer is measured by absorption imaging and normalized to the initial total number of atoms regardless of spin state.

\begin{figure}[ht!]
\includegraphics[width=1\linewidth]{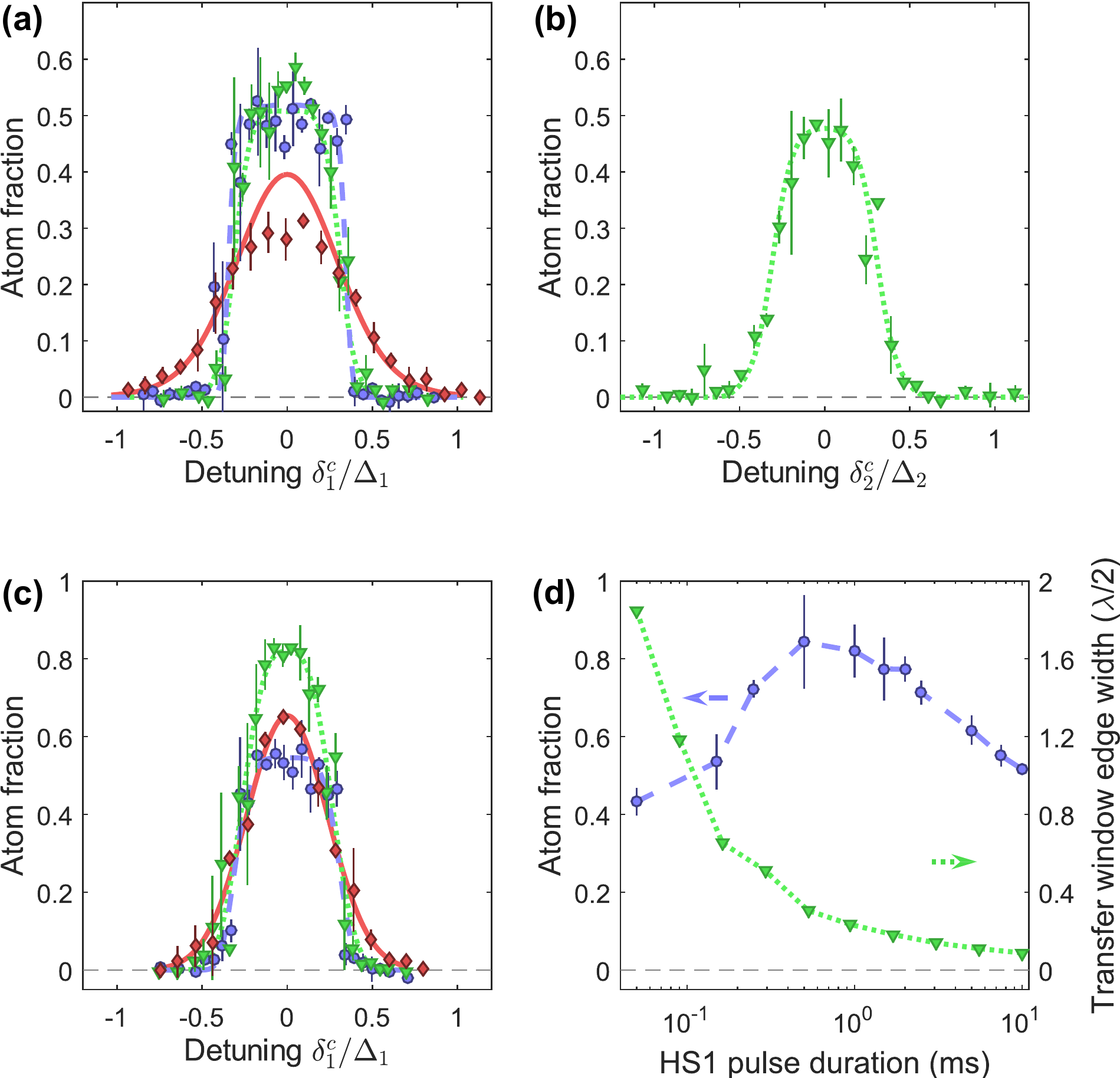}
\caption{Characterization of the microwave transfer for an equal mixture of spin states in a homogeneous magnetic field. \textbf{(a)} Transfer of atoms in state $\ket{\uparrow}$ with pulse width $\delta^0_1=7\,\mathrm{kHz}$ and pulse durations $T=0.15\,\mathrm{ms}$ (red diamonds and solid line), $1\,\mathrm{ms}$ (green triangles and dotted line), $10\,\mathrm{ms}$ (blue circles and dashed line). The lines are a fit to a numerical model (see text). \textbf{(b)} Same as (a) for the transfer of state $\ket{\downarrow}$ with pulse width $\delta^0_2=5.25\,\mathrm{kHz}$ and a pulse duration of $1\,\mathrm{ms}$. \textbf{(c)} Transfer of atoms from the initial spin mixture, with pulse widths $\delta^0_1=7.0\,\mathrm{kHz}, \delta^0_2=5.25\,\mathrm{kHz}$ and pulse durations $T=0.15\,\mathrm{ms}$ (red diamonds and solid line), $1\,\mathrm{ms}$ (green triangles and dotted line), $10\,\mathrm{ms}$ (blue circles and dashed line). The lines are a guide for the eye. \textbf{(d)} Comparison for a variable duration of the measured transfer efficiency for a mixture of states $\ket{\uparrow}$ and $\ket{\downarrow}$ (blue circles and dashed line) and of the calculated sharpness of the edge of the transfer pulse in frequency space (green triangles and dotted line), expressed as the corresponding distance in the magnetic field gradient for transition 1 in units of the lattice spacing.}
\label{pulses}
\end{figure}

Figure \ref{pulses}(a) shows the transfer profile of the microwave pulses for the preparation of atoms in state $\ket{\uparrow}$ using the steps B, C and D of our protocol. We observe a clear dependence of the transfer efficiency on the center frequency of the microwave pulse. Frequency sweeps which do not cross the resonant transition frequency do not transfer atoms to the $F=7/2$ manifold, and cause a complete removal of the atoms by the laser pulse C. A maximal transfer is obtained for frequency sweeps that are centered at resonance, with a transfer efficiency dependent on the pulse duration $T$. If we satisfy the adiabaticity condition ($T\ge1$\,ms) the transfer efficiency is close to 100\% in a frequency range of approximately $\delta^0_i$, leading to a flat-top spectral transfer profile. Shorter pulse durations, e.g. for $T=0.15\,\mathrm{ms}$ in figure \,\ref{pulses}(a), violate the adiabaticity condition and lead to a reduced transfer efficiency with less sharp edges.

The datasets (four curves on figures \ref{pulses}(a) and (b)) are simultaneously fitted with our numerical calculation of the transfer profiles, knowing the total atom number in the sample from a separate measurement, with the fraction of atoms in the $\ket{\uparrow}$ state in the initial sample as the only fitting parameter. We find good agreement between the theoretical curves and experimental data for a fraction of atoms in the $\ket{\uparrow}$ state of 52(1)\%. In particular, the fact that no other adjustable parameter is needed means that the efficiency of our transfer pulses is close to its theoretical value, which is 98\% for a $T=1\,\mathrm{ms}$ pulse and for the \ket{\uparrow} state.
A deviation between the measured data and the theoretical curve is observed for a pulse duration of $T=150\,\mathrm{\mu s}$, which can be attributed to small imperfections of the pulse shape, that become critical in the regime where adiabaticity is not satisfied.

In a final step we combine the microwave pulses A, E and B, D for the transfer of both spin states (figure \,\ref{pulses}(c)). Similarly to the transfer of one spin state, we observe an increase of the total transfer efficiency up to 82(3)\% if we increase $T$ from 0.15\,ms to 1\,ms. Surprisingly, there is a reduction of the transfer efficiency to 58(2)\% for a longer pulse duration of 10\,ms (blue circles in figure\,\ref{pulses}(b)). It is caused by collisional loss in the $F=7/2$ manifold, when we transiently prepare atoms in a mixture of the $\ket{7/2,-7/2}$ and $\ket{7/2,-5/2}$ states. This loss is investigated in detail in Section \ref{sec:losses}. Figure \,\ref{pulses}(d) demonstrates the compromise that needs to be made between adiabaticity (transfer is inefficient for fast pulses) and loss (the number of transferred atoms drops for longer durations). At the same time, a longer pulse duration provides a sharper spectral profile, minimizing the transfer in layers other than $\mathcal{L}_0$. Consequently, we use a transfer pulse duration of $T=1\,\mathrm{ms}$, as it gives a flat-top spectral profile with 98\% transfer efficiency for the transfer of individual spins.
This flat-top transfer profile over a frequency range of about $0.25\Delta_i$ (2.5\,kHz for state  \ket{\uparrow}) ensures that the preparation is robust against small changes of the resonance frequency. From the sharp edges of the spectra in figure\,\ref{pulses}(a), we estimate frequency fluctuations of less than 1\,kHz for transition 1. This shows that the transition is stable at this level on timescales of $<30$\,min, which is the time it takes to acquire a data set like those presented in figure\,\ref{pulses}(a). We found that over a full day, the transition frequency can drift by up to 5\,kHz, due to changes of the ambient magnetic field or due to temperature changes of the setup which shift the position of the magnetic coils with respect to the atoms. This change in frequency requires us to re-center our transfer window on a layer every few hours.

\subsection{Loss in the F=7/2 manifold}
\label{sec:losses}

\begin{figure}[ht!]
\includegraphics[width=0.9\linewidth]{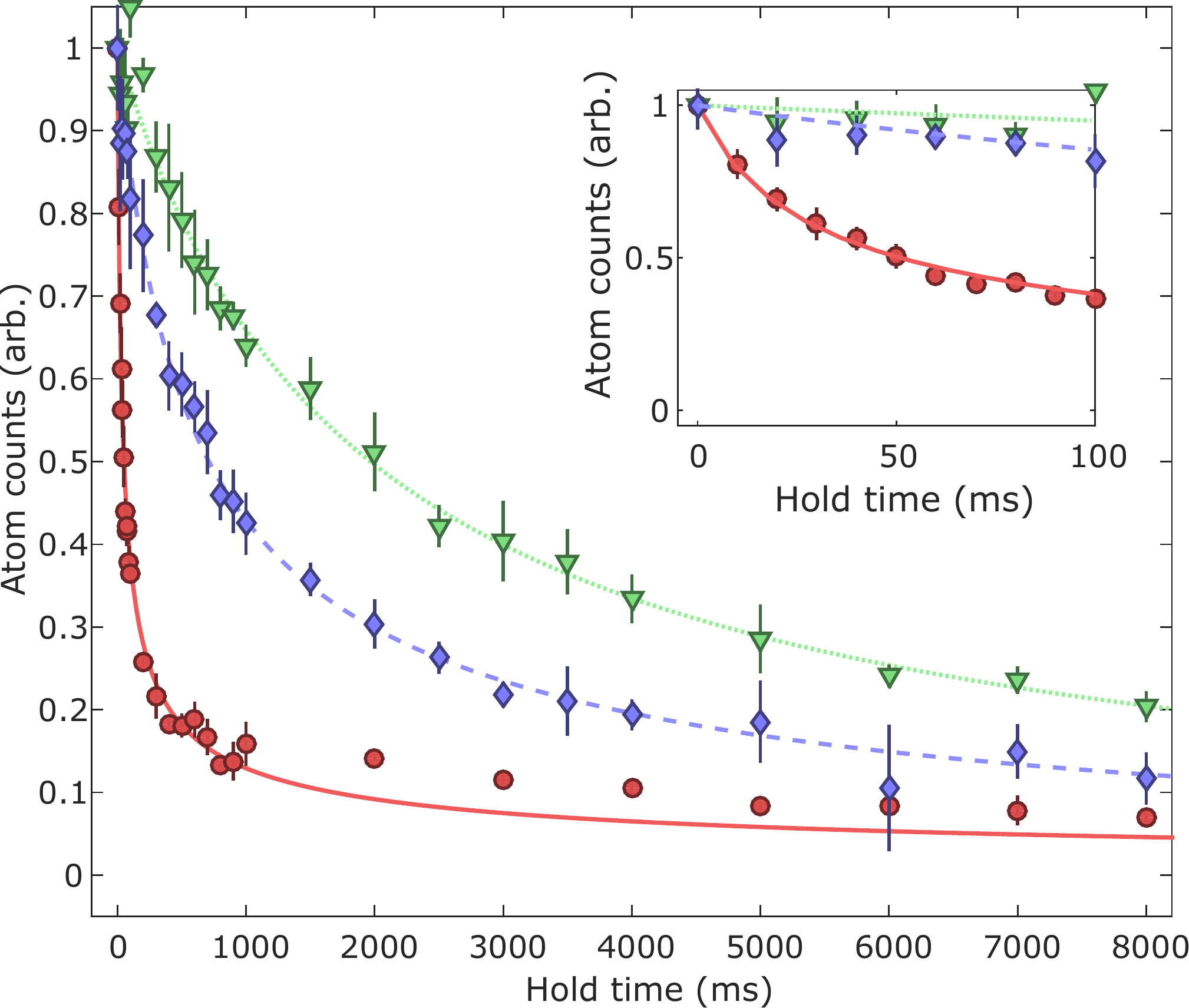}
\caption{Atom loss in the $\ket{F=7/2,m_F=-7/2}$ state (green triangles and dotted line), $\ket{F=7/2,m_F=-5/2}$ state (blue diamonds and dashed line), or an equal-weight mixture of the two (red circles and solid line). The atom count is normalized to its initial value for readability. Each dataset is fitted with a decay curve obtained from numerical integration of the rate equation \ref{rateeq} accounting for two-body and three-body losses (see text for details).}
\label{losses}
\end{figure}

We observe loss of atoms during the single-layer preparation of states in the $F=7/2$ manifold. It occurs on a timescale of 100\,ms which is significantly shorter than our measured lifetime of about 30\,s for a balanced mixture of states $\ket{\uparrow}$ and $\ket{\downarrow}$. A possible reason for this are hyperfine-state-changing collisions in the $F=7/2$ manifold, which release sufficient energy to lead to trap loss \cite{mudrich2004}.

A measurement of the time-dependence of the atom loss for individual spin states and in a state mixture in the $F=7/2$ manifold is shown in figure \,\ref{losses}. We use the steps A and E of our preparation procedure to transfer $2.3\times 10^4$ atoms to the state $\ket{m_F=-7/2}$ (green triangles), the steps B and C to transfer about $2\times 10^4$ atoms in the state $\ket{m_F=-5/2}$ (blue diamonds), and A, B, and C to prepare $3.5\times 10^4$ atoms in an equal mixture of both states (red circles). All measurements were performed in a 44.4(5) $E_r$ depth optical lattice, corresponding to trapping frequencies $\omega_z/(2\pi)=58.8(3)\,\mathrm{kHz}$ (along the lattice axis) and $\omega_{x,y}/(2\pi)=128(1)\,\mathrm{Hz}$. The initial temperature of our samples is approximately $0.5\,\mathrm{\mu K}$.

Half of the atoms are lost after a duration of 2000\,ms, 700\,ms and 50\,ms for atoms in the states $\ket{m_F=-7/2}$, $\ket{m_F=-5/2}$ and in a spin mixture, respectively. Each dataset is fitted by a numerical solution to a rate equation including two-body and three-body loss terms (and omitting trap lifetime over the timescales considered):
\begin{eqnarray}
\label{rateeq}
\frac{\mathrm{d}N}{\mathrm{d}t}= - \alpha N^2 - \beta N^3
\end{eqnarray}
where $N$ is the atom number for the sample considered, and $\alpha$ and $\beta$ characterize two- and three-body losses, respectively.

Loss in the $F=7/2$ manifold occur much faster than for atoms in the $F=9/2$ manifold, even when the sample is prepared in a single quantum state, when $s-$wave collisions should be suppressed due to the fermionic nature of the atoms. This may indicate the presence of $p-$wave collisions between atoms in the same magnetic state.
Collisions in a mixture of states $\ket{m_F=-7/2}$ and $\ket{m_F=-5/2}$ cause atom loss on an even shorter timescale of a few milliseconds. Our fit to the data provides the following values for the decay constants: $\alpha=0.4(2)\times 10^{-3}\,\mathrm{s^{-1}}$ (resp. $\alpha=6(3)\times 10^{-4}\,\mathrm{s^{-1}}$ and $\alpha=0(2)\times 10^{-3}\,\mathrm{s^{-1}}$) and  $\beta=0.1(2)\,\mathrm{s^{-1}}$ (resp. $\beta=1.2(5)\times 10^{-3}\,\mathrm{s^{-1}}$ and $\beta=2.9(4)\times 10^{-2}\,\mathrm{s^{-1}}$) for atoms in state $\ket{m_F=-7/2}$ (resp. $\ket{m_F=-5/2}$ and an equal mixture of the two).
The simple model of equation \ref{rateeq} does not take into account the possible change of temperature of the sample \cite{weber2003}, which may be responsible for the imperfect fit of the data for the spin mixture.
A full analysis of these loss mechanisms in the $F=7/2$ manifold, dependent on the density distribution of the atoms, including trap geometry and temperature, is beyond the scope of this publication.

These measurements eventually lead us to choose a pulse duration of $T=1$\,ms that is a compromise between satisfying the adiabaticity condition (long pulse durations) and minimizing the atom loss during preparation (short pulse durations).

\section{\label{sec:gradients}Imaging magnetic field inhomogeneities}

The square transfer window of the adiabatic passages makes the single-layer preparation process robust against small frequency drifts. In addition, we can exploit these sharp spectral features to detect and visualize small magnetic field gradients. We freeze the position of the atoms in a deep 3D optical lattice potential and apply our preparation scheme to one spin state ($\ket{\uparrow}$). Only atoms which experience a Zeeman shift that matches the microwave transition frequencies of the narrow transfer window are transferred to the $F=7/2$ manifold, and are not affected by the optical removal pulse. By imaging the spatial distribution of remaining atoms with fluorescence imaging, we can infer the distribution of the transition frequencies and the magnetic field strengths, a technique referred to as spectral imaging \cite{Marti2018}.

Large vertical magnetic field gradients are essential for our scheme to select a single layer of the vertical lattice potential, but horizontal magnetic field gradients are detrimental because they result in a transition frequency spread across the layer. 
We expect the quadrupole field generating the vertical field gradient to induce a horizontal field curvature. Ideally, the vertical axis of the quadrupole field is centered on the atoms, and the surfaces of equal magnetic field magnitude are ellipsoids with a zero crossing of the horizontal field gradient on the axis. However, external stray magnetic fields can shift the vertical axis of the quadrupole field, and transverse gradients may also stem from possible misalignments or tilts of the coils in the setup. 

\begin{figure}[t!]
\includegraphics[width=\linewidth]{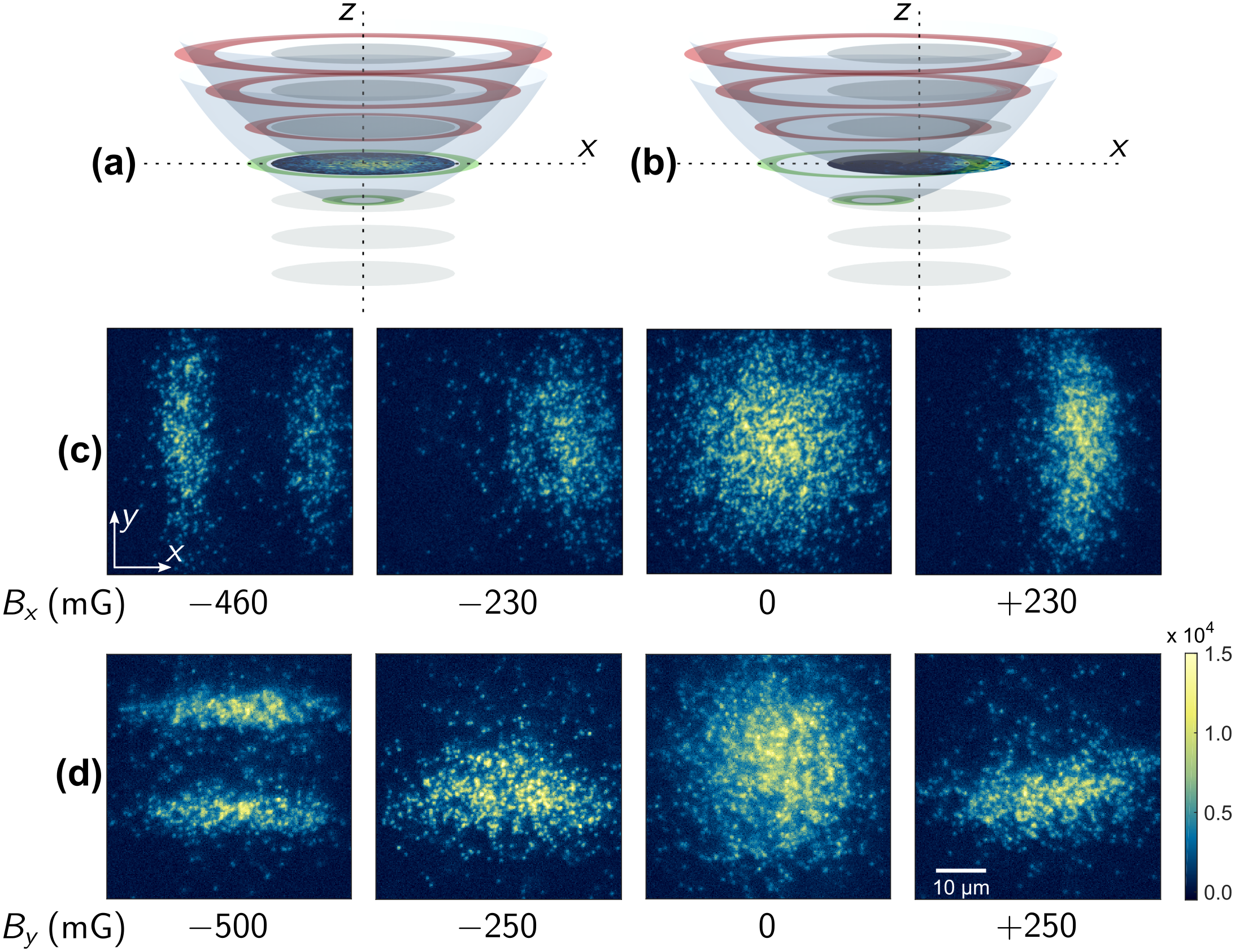}
\caption{Compensation of stray magnetic fields in two directions in the imaging plane of the microscope objective (see text for details). \textbf{(a)} and \textbf{(b)} Depiction of the surfaces of equal transition energy for a horizontally centered quadrupole field (a) or out-of-center (b), corresponding to the two last images of (c). \textbf{(c)} Varying compensation magnetic field along $x$. \textbf{(d)} Varying compensation field along $y$.}
\label{gradients}
\end{figure}

Figure\,\ref{gradients} illustrates the spatial selection of atoms for a well-centered (a) and a shifted (b) quadrupole field axis. The transfer window defines the range of transition frequencies of addressable atoms. A given transition frequency defines an ellipsoid with constant magnetic field strengths as indicated in figure\,\ref{gradients}(a) by curved surfaces. As a result, the position of the addressed atoms is within the intersection of the initially occupied 3D lattice sites and the ellipsoids, with a width set by the size of the transfer window. Red and green surfaces in figure \ref{gradients}(a) indicate possible geometric shapes of the selected atoms, such as discs, rings and curved stripes, for two center frequencies of the transfer window. In figure \ref{gradients}(a), the center frequency that allows to address atoms at the focal plane of the microscope corresponds to the red surfaces. In figure \ref{gradients}(b), with a  displacement of the quadrupole, a different ellipsoid (with green surfaces) intersects with the atom sample at the focus of the microscope, producing a curved stripe pattern on the fluorescence image.

In order to image and compensate the existing transverse gradients, we prepare a balanced mixture of spin states in a 3D optical lattice of depth $44.4(5)\,E_r$ along the vertical axis and $38.4(4)\,E_r$ along horizontal axes. The atoms are thus at a fixed position when we selectively prepare them in the $\ket{\uparrow}$ state (applying steps B, C and D of our protocol, see figure \ref{setup}(b)), using a narrow microwave sweep width of $\delta^0_{1}=3.0\,\mathrm{kHz}$. These atoms are then detected by fluorescence imaging  \cite{Haller2015}. 

To visualize the effect of magnetic field gradients, we employ our horizontal shim coils (figure\,\ref{setup}(a)) to intentionally displace the minimum of the quadrupole field in the horizontal plane. The effect on our fluorescence images is clearly visible in figure\,\ref{gradients}. Depending on the shim fields applied, the images show various striped patterns as described above. We can use this spectral imaging scheme to measure and minimize horizontal magnetic field gradients with high precision. By increasing the spacing of the stripe pattern, we effectively reduce the magnetic field gradient until a single stripe covers the full field of view. The scheme can be applied in both horizontal directions (see figure\,\ref{gradients}(c) and (d)).

The residual magnetic field inhomogeneities across a single layer can be estimated using the frequency widths of the microwave pulses ($\delta^0_{1}=3.0\,\mathrm{kHz}$) and our field of view ($50\,\mathrm{\mu m}$). Our spectral imaging scheme therefore allows us to reduce horizontal magnetic field inhomogeneities to less than $1.2\,\mathrm{mG}$. This value is compatible with our estimate of the maximum variation of magnetic field due to the quadrupole field when it is centered, of about $0.15\,\mathrm{mG}$ across the imaging plane.

\section{Conclusion}

Quantum-gas microscopes rely on the fluorescence imaging of ultracold atomic gases in a 2D geometry in the focal plane of a high-resolution microscope objective. The imaging process requires the preparation of a quantum system in a single layer of an optical lattice potential. In this article, we demonstrated such a preparation scheme for a spin-mixture of fermionic potassium atoms. Our technique relies on a transfer of the atoms to other spin states, and the spatial selection of the atoms is achieved by adiabatic microwave pulses in a magnetic field gradient. Our method is applicable to mixtures of other atomic species with a non-zero magnetic moment.
To characterize and optimize our single-layer preparation scheme, we studied loss that occurs for atoms in spin-states of the $F=7/2$ manifold. We also demonstrated that our frequency-selective preparation protocol can be used to image and compensate magnetic gradients. Our protocol can be transferred to other atomic species and it will be useful to facilitate the study of many-body quantum systems in optical lattices in lower-dimensional systems.

\bigskip
\paragraph*{\bf Acknowledgments:}
We acknowledge financial support from the H2020 EU collaborative project ``QuProCS'' (Grant Agreement 641277) and from the EPSRC Programme Grant ``DesOEQ'' (Grant Agreement EP/P009565/1). The data and source code used in this publication are openly available from the University of Strathclyde KnowledgeBase in \cite{metadata}.

\bibliography{Peaudecerf2018}

\end{document}